\documentclass[prl,twocolumn,floatfix,showpacs]{revtex4}

\usepackage{amssymb}
\usepackage{amsmath}
\usepackage{amsfonts}
\usepackage{graphicx}

\newtheorem{theorem}{Theorem}

\newenvironment{proof}[1][Proof]{\noindent\textbf{#1.} }{\ \rule{0.5em}{0.5em}}

\begin{document}

\title{Quantum phase transitions and bipartite entanglement}
\author{L.-A. Wu, M. S. Sarandy, D. A. Lidar}
\affiliation{Chemical Physics Theory Group, Department of Chemistry, and Center for
Quantum Information and Quantum Control, University of Toronto, 80 St.
George St., Toronto, Ontario, M5S 3H6, Canada}

\begin{abstract}
We develop a general theory of the relation between quantum phase
transitions (QPTs) characterized by nonanalyticities in the energy and bipartite entanglement. 
We derive a functional relation between the matrix elements of two-particle reduced density 
matrices and the eigenvalues of general two-body Hamiltonians of $d$-level
systems. The ground state energy eigenvalue and its derivatives, whose
non-analyticity characterizes a QPT, are directly tied to bipartite
entanglement measures. We show that first-order QPTs are
signalled by density matrix elements themselves and second-order QPTs by the first
derivative of density matrix elements. Our general conclusions are
illustrated via several quantum spin models.
\end{abstract}


\maketitle

Recently, a great deal of effort has been devoted
to the understanding of the connections between quantum information
\cite{Chuang:book} and the theory of quantum critical phenomena
\cite{Sachdev:book}. A key novel observation is that quantum
entanglement can play an 
important role in a quantum phase transition (QPT) \cite{Osterloh:02,Osborne:02,JVidal:04a,Huang:04,
Vidal:03,Vidal:04,Verstraete:04a,Barnum:04,Somma:04,Bose:02,Alcaraz:03,JVidal:04,Gu:03,Lambert:04}. 
In particular,
for a number of spin systems, it has been shown that QPTs are signalled by a 
critical behavior of bipartite entanglement as measured, for instance, in terms of
the concurrence~\cite{Wootters:98}. For the case of second-order QPTs (2QPTs), the critical point was found 
to be associated to a singularity in the derivative of the ground state concurrence, as
first illustrated, for the transverse field Ising chain, in
Ref.~\cite{Osterloh:02}, and generalized in 
Refs.~\cite{Osborne:02,JVidal:04a,Huang:04}
(see Refs. \cite{Verstraete:04a,Barnum:04,Somma:04,Vidal:03,Vidal:04} for an analysis in terms of other
entanglement measures). In the case of first-order QPTs (1QPTs), discontinuities
in the ground state concurrence were shown to detect
the QPT \cite{Bose:02,Alcaraz:03,JVidal:04}. The studies conducted to
date are based on the analysis of particular many-body models. Hence the
general connection between bipartite entanglement and QPTs is not yet well
understood. The aim of this work is to discuss, in a general framework, how
bipartite entanglement can be related to a QPT characterized by nonanalyticities in the energy.

\textit{Expectation values and the reduced density matrix}.--- The most
general Hamiltonian of non-identical
particles, up to two-body interactions, reads 
\begin{equation}
H=\sum_{i\alpha \beta }\epsilon _{\alpha \beta }^{i}\left\vert \alpha
_{i}\right\rangle \left\langle \beta _{i}\right\vert +\sum_{ij\alpha \beta
\gamma \delta }V_{\alpha \beta \gamma \delta }^{ij}\left\vert \alpha
_{i}\right\rangle \left\vert \beta _{j}\right\rangle \left\langle \gamma
_{i}\right\vert \left\langle \delta _{j}\right\vert ,
\label{eq1}
\end{equation}
where $\{\left\vert \alpha _{i}\right\rangle \}$ is a basis for the
Hilbert space, $\alpha ,\beta,\gamma ,\delta \in \{0,1,...,d-1\}$, and $i,j$ enumerate $N$ ``qudits''
($d$-level systems). Let $E=\langle \psi |H|\psi \rangle $ be the energy in a non-degenerate eigenstate 
$\left\vert \psi \right\rangle $ of the Hamiltonian. The two-spin reduced density operator 
${\hat{\rho}}^{ij}$ is given by ${\hat{\rho}}^{ij}=\sum_{m}\left\langle m|\psi \right\rangle \left\langle
\psi |m\right\rangle$, with $m$ running over all the $d^{N-2}$ orthonormal basis vectors, excluding
qudits $i$ and $j$. ${\hat{\rho}}^{ij}$ has a $d^{2}\times d^{2}$ matrix representation ${\rho }^{ij}$, with elements
${\rho }_{\gamma \delta \alpha \beta }^{ij} =\left\langle \gamma _{i}\delta_{j}\right\vert {\hat{\rho}}^{ij}\left\vert \alpha _{i}\beta
_{j}\right\rangle =\sum_{m}\left\langle \gamma _{i}\delta _{j}m|\psi\right\rangle \left\langle \psi |m\alpha _{i}\beta _{j}\right\rangle
=\sum_{m}\left\langle \psi |m\alpha _{i}\beta _{j}\right\rangle\left\langle \gamma _{i}\delta _{j}m|\psi \right\rangle =\left\langle \psi
|\alpha _{i}\beta _{j}\right\rangle \left\langle \gamma _{i}\delta _{j}|\psi\right\rangle$, 
where we have used that $\left\langle \psi |m\alpha _{i}\beta_{j}\right\rangle $ are $c$-numbers and $\sum_{m}\left\vert m\right\rangle
\left\langle m\right\vert =1$. Similarly, we can show that ${\hat{\rho}}^{i}=\mathrm{Tr}_{j}({\hat{\rho}}^{ij})$ 
has a $d\times d$ matrix representation ${\rho }^{i}$ with elements 
$\rho _{\beta \alpha }^{i}=\left\langle \beta _{i}\right\vert {\hat{\rho}}
^{i}\left\vert \alpha _{i}\right\rangle =\left\langle \psi |\alpha
_{i}\right\rangle \left\langle \beta _{i}|\psi \right\rangle$.
Therefore, the energy $\langle \psi |H|\psi \rangle$ is 
\begin{equation}
E(\rho ^{ij}) = \sum_{ij}\mathrm{Tr}(\mathbf{U}(ij){\rho }^{ij}\mathbf{)},  \label{erho}
\end{equation}
with $\mathbf{U}(ij)$ denoting a $d^{2}\times d^{2}$ matrix whose elements are 
$U_{\alpha \beta ,\gamma \delta }(ij)=\epsilon _{\alpha \gamma }^{i}\delta
_{\beta \delta }^{j}/N_i+V_{\alpha \beta \gamma \delta }^{ij}$,
where $N_i$ is the number of qudits that qudit $i$ interacts with, and $\delta _{\beta \delta }^{j}$ 
is the Kronecker symbol on qudit $j$. Clearly, Eq.~(\ref{erho}) holds not only for the 
Hamiltonian operator but for any observable. Indeed, it turns out that the expectation value (or
eigenvalue, for an eigenstate) of any two-qudit observable in an arbitrary 
state $\left\vert \psi \right\rangle$ is a linear
function of the matrix elements of two-spin reduced density matrices.
Moreover, it is easy to show that Eq.~(\ref{erho}) is also valid for a set of qudits 
with distinct dimensions and for an
arbitrary $D$-fold degenerate energy level, where $\rho^{ij} = (1/D)\sum_{p=1}^D \rho^{ij}_p$, 
with $\rho^{ij}_p$ denoting the reduced density operator 
associated to the degenerate eigenstate $|\psi_p\rangle$. These
results easily generalize to the case
of a Hamiltonian containing $n$-body terms; e.g., for a three-body
operator $\hat{O}$, $\left\langle \psi \right\vert \hat{O}\left\vert \psi
\right\rangle
=\sum_{ijk}\mathrm{Tr}(\mathbf{O}(ijk){\rho}^{ijk})$, etc., for
higher-order interactions.
The above results hold for any value of $d$. Here we are especially
interested in $d=2$, i.e., the qubit case. We then use the standard basis $
\left\{ \left\vert 00\right\rangle ,\left\vert 01\right\rangle ,\left\vert
10\right\rangle ,\left\vert 11\right\rangle \right\} $ for any pair $(i,j)$
of spins, and denote $\rho _{11}^{ij}=\left\langle
0_{i}0_{j}\right\vert {\hat{\rho}}^{ij}\left\vert
0_{i}0_{j}\right\rangle $, $\rho _{12}^{ij}=\left\langle
0_{i}0_{j}\right\vert {\hat{\rho}}^{ij}\left\vert
0_{i}1_{j}\right\rangle $, etc.

\textit{QPT and the reduced density operator}.--- QPTs are critical changes
in the properties of the ground state of a many-body system due to
modifications in the interactions among its constituents, occurring at low
temperatures $T$ where the de Broglie thermal wavelength is greater than the
classical correlation length of the thermal fluctuations (effectively $T=0$) 
\cite{Sachdev:book}. Typically, such a change is induced as a parameter $
\lambda $ in the system Hamiltonian $H(\lambda )$ is varied across a
critical point $\lambda _{c}$. Because they occur at $T=0$, QPTs are purely
driven by quantum fluctuations. They are associated with level crossings which, 
in many cases, lead to the presence of nonanalyticities in the energy spectrum. 
Specifically, a 1QPT is characterized by a finite discontinuity in the first derivative 
of the ground state energy. A 2QPT (or continuous QPT) is similarly characterized by a
finite discontinuity, or divergence, in the second derivative of the ground
state energy, assuming the first derivative is continuous. These
characterizations are the $T=0$ limits of the classical definition of the
corresponding phase transitions, given in terms of the free energy~\cite{Reichl}.
There are QPTs where this is not the case~\cite{Zanardi:2004,Yang:04}. One such example is 
the QPT in the antiferromagnetic XXZ model, where a critical anisotropy separates a gapless phase 
from a gapful phase. As shown in Ref.~\cite{Yang:66}, the ground state energy and all of its derivatives with 
respect to the anisotropy are continuous at the critical point, despite the existence of the QPT. 
Moreover, other examples where QPTs are not directly related to nonanlyticities in the ground 
state energy include transitions caused by level crossings in the low-lying excited states~\cite{Nomura:94,Tian:03} and 
those associated with topological order (e.g., in fractional quantum Hall liquids), which is not
characterized by symmetry breaking~\cite{Wen:04}. We shall consider in this Letter only QPTs characterized by nonanalytic behavior 
in the derivatives of the ground state energy. 

Assume that $\epsilon _{\alpha \beta }^{i}$ and $V_{\alpha \beta \gamma
\delta }^{ij}$ are smooth functions of a set $\left\{ \lambda _{k}\right\} $
of couplings. If $\left\vert \psi \right\rangle $ is an eigenstate of the
Hamiltonian then, using $\partial_\lambda \langle \psi \left\vert \psi
\right\rangle =0 \Rightarrow  \partial_\lambda E
=\left\langle \psi \right\vert \partial_\lambda H \left\vert \psi
\right\rangle $, we have from Eq.~(\ref{erho}):
\begin{equation}
\partial_\lambda \mathcal{E}(\rho ^{ij}) = (1/N)
\sum_{ij}\mathrm{Tr}([\partial_\lambda \mathbf{U}(ij)]\rho^{ij}),
\label{fd}
\end{equation}
where $\mathcal{E}=E/N$. It follows immediately from Eq.~(\ref{fd}) that 
$\sum_{ij}\mathrm{Tr}(\mathbf{U}(ij)[\partial_\lambda \rho ^{ij}])=0$.
The origin of a 1QPT can now be seen to be the discontinuity of
one or more of the $\rho ^{ij}$'s at the critical point. The second
derivative, obtained directly from Eq.~(\ref{fd}), reads 
\begin{equation}
\frac{\partial ^{2}\mathcal{E}(\rho ^{ij})}{\partial \lambda ^{2}}=\frac{1}{N
}\sum_{ij}\left\{ \mathrm{Tr}(\frac{\partial ^{2}\mathbf{U}(ij)}{\partial
\lambda ^{2}}\rho ^{ij})+\mathrm{Tr}(\frac{\partial \mathbf{U}(ij)}{\partial
\lambda }\frac{\partial \rho ^{ij}}{\partial \lambda })\right\}  \label{sd}
\end{equation}
Since $\mathbf{U}(ij)$ is a smooth function of $\left\{ \lambda _{k}\right\} 
$ and $\rho ^{ij}$ is finite at the critical point $\lambda =\lambda _{c},$
it can now similarly be seen that the origin of the discontinuity or
singularity of $\partial ^{2}\mathcal{E}(\rho ^{ij})/\partial \lambda ^{2}$
is due to the fact that one or more of the $\partial \rho ^{ij}/\partial
\lambda $'s diverge at the critical point.

\textit{QPTs from bipartite entanglement}.--- In order to discuss the role
of bipartite entanglement in a QPT we need appropriate entanglement measures 
$M(\rho ^{ij})$: monotonic functions ranging from $0$ (no entanglement) to $
1$ (maximal entanglement), invariant under local operations and classical
communication \cite{Chuang:book}. We consider two such measures: 
(i) \emph{concurrence} \cite{Wootters:98}: $C(\rho ^{ij})=\max ( \gamma _{1}^{ij}-\gamma _{2}^{ij}-\gamma
_{3}^{ij}-\gamma _{4}^{ij},0)$,
where the $\gamma _{\alpha }^{ij}$ are the square-roots, in decreasing
order, of the eigenvalues of the operator $R(\rho ^{ij})\equiv \rho
^{ij}(\sigma _{y}\otimes \sigma _{y})\rho ^{ij\ast }(\sigma _{y}\otimes
\sigma _{y})$, where $\rho ^{ij\ast }$ denotes complex conjugation of $\rho
^{ij}$ in the standard basis $\left\{ \left\vert 00\right\rangle ,\left\vert
01\right\rangle ,\left\vert 10\right\rangle ,\left\vert 11\right\rangle
\right\} $; (ii) \emph{negativity} \cite{Vidal:02a}:
$\mathcal{N}(\rho ^{ij})=2\,\max (0,-\min_{\alpha }(\mu _{\alpha }^{ij}))$,
where $\mu _{\alpha }^{ij}$ are the eigenvalues of the partial transpose $
\rho ^{ij,T_{A}}$ of the density operator $\rho ^{ij}$, defined as $
\left\langle \alpha \beta \right\vert \rho ^{T_{A}}\left\vert \gamma \delta
\right\rangle =\left\langle \gamma \beta \right\vert \rho \left\vert \alpha
\delta \right\rangle $.

It is now a simple matter to connect these measures to the appearance of a
QPT. From Eq.~(\ref{erho}) we have 
$E(\rho ^{ij})=\sum_{ij}\mathrm{Tr}(\mathbf{U}(ij)\rho ^{ij})=
\sum_{ij}\mathrm{Tr}(\mathbf{U}^{T_{A}}(ij)\rho ^{ij,T_{A}})$,
where the matrix elements of $\mathbf{U}^{T_{A}}(ij)$ are $\left\langle
\alpha \beta \right\vert \mathbf{U}^{T_{A}}\left\vert \gamma \delta
\right\rangle =\left\langle \gamma \beta \right\vert \mathbf{U}\left\vert
\alpha \delta \right\rangle $. Let $W^{ij}$ be the unitary matrix that
diagonalizes $\rho ^{ij,T_{A}}$. Then, using Eq.~(\ref{fd}), we obtain 
\begin{equation}
\frac{\partial \mathcal{E}(\rho ^{ij})}{\partial \lambda }=\frac{1}{N}
\sum_{ij}\sum_{\alpha =1}^{4}\left\{ W^{ij}\frac{\partial \mathbf{U}^{T_{A}}(ij)}{
  \partial \lambda }W^{ij\dagger }\right\} _{\alpha \alpha }\mu
_{\alpha}^{ij}.
\label{neg}
\end{equation}

\begin{theorem}
Assume conditions (a)-(c) below are satisfied. Then: a discontinuity in
[discontinuity in or divergence of the first derivative of] the
concurrence or negativity is both necessary and sufficient to signal a
1QPT [2QPT]. 

\vspace{0.1cm}

\noindent (a) The 1QPT [2QPT] is associated to a discontinuity in [discontinuity in or
divergence of] the first [second] derivative of the ground state energy, which
originates \emph{exclusively} from the elements of $\rho^{ij}$ and not, for
instance, from the sum in Eq.~(\ref{fd}) [Eq.~(\ref{sd})] itself. Similarly, a
discontinuity in [discontinuity in or divergence of the first derivative of]
the concurrence or negativity originates \emph{exclusively} from $\rho^{ij}$
and not from other operations such as max or min. 

\vspace{0.1cm}

\noindent (b) In the case of a 1QPT [2QPT] the discontinuous matrix
elements of $\rho^{ij}$ present in Eq.~(\ref{fd}) [discontinuous or
divergent $\partial\rho^{ij}/\partial \lambda$ present in Eq.~(\ref{sd})] 
do not either all accidentally vanish or cancel with other terms in the 
expression for [the first derivative of] the concurrence or negativity.

\vspace{0.1cm}

\noindent (c) In the case of a 1QPT [2QPT] the discontinuous matrix elements of $\rho^{ij}$ 
present in [discontinuous or divergent $\partial\rho^{ij}/\partial \lambda$ 
present in the first derivative of] the concurrence 
or negativity do not either all accidentally vanish or cancel with other terms in 
Eq.~(\ref{fd}) [Eq.~(\ref{sd})].

\end{theorem}

Conditions (a)-(c) above are meant to exclude artificial/accidental
occurrences of non-analyticity. They are meant to emphasize that 
the entanglement-QPT connection may directly come from the
ground state reduced density matrix. When non-analyticities originating from the density operator are present
in both the entanglement measure (or its derivatives) and the derivatives of the ground state energy,
bipartite entanglement and QPTs signal each other. These observations are also the basis of the
proof we now give.

\vspace{0.1cm}

\begin{proof}
\emph{1QPT}: If condition (a) is satisfied then a 1QPT must come 
from the discontinuity of one (or more) matrix elements of $\rho^{ij}$, as given 
by Eq.~(\ref{fd}). Thus, 
taking into account condition (b), the 1QPT will 
be associated to a discontinuity in the concurrence or negativity, which is therefore 
a necessary condition for the 1QPT. Sufficiency: (i) Concurrence --
Taking into account condition (a), if one (or more) of
the eigenvalues $\gamma _{\alpha }^{ij}$ of $R(\rho ^{ij})$ is discontinuous then one (or more) of the
matrix elements of $\rho ^{ij}$ must be discontinuous. Assuming condition (c),  
a 1QPT then follows from Eq.~(\ref{fd}). 
(ii) Negativity -- the negativity and $\partial \mathcal{E}(\rho
^{ij})/\partial \lambda $ are both linear in $\min_{\alpha }(\mu _{\alpha}^{ij})$. Therefore if the coefficient in front of $\min_{\alpha }(\mu
_{\alpha }^{ij})$ in Eq.~(\ref{neg}) does not accidentally vanish, as ensured  
by condition (c), a discontinuous negativity signals the 1QPT. 

\emph{2QPT}:
Considering Eq.~(\ref{sd}), if condition (a) is satisfied then a 
2QPT must come from the discontinuity in or divergence of one (or
more) 
$\partial\rho^{ij}/\partial \lambda$, since all the $\rho^{ij}$ are
assumed to be continuous for the case of a 2QPT. Thus, taking into
account condition (b),  
the 2QPT will be associated to a discontinuity in or divergence of the first derivative 
of the concurrence or negativity, which is therefore a necessary condition for the 2QPT. On the 
other hand, we have $
\partial_\lambda M(\rho ^{ij}) = \sum_{a,b=1}^{4}
[\partial M(\rho ^{ij})/\partial \rho _{ab}^{ij}] \partial_\lambda \rho_{ab}^{ij}$.  
Therefore, taking into account condition (a), discontinuity in or divergence of $[\partial M(\rho ^{ij})/\partial
\lambda ]_{\lambda _{c}}$ must be caused by one or more of the $
[\partial \rho _{ab}^{ij}/\partial \lambda ]_{\lambda _{c}}$. Assuming condition  
(c), this singular behavior of $\partial\rho^{ij}/\partial \lambda$ is then 
a sufficient condition for a 2QPT, which follows from Eq.~(\ref{sd}). 
\end{proof}

\begin{figure}[ht]
\centering {\includegraphics[angle=0,scale=0.29]{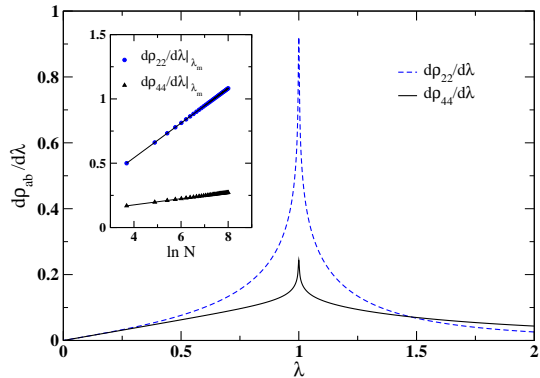}}
\caption{First derivative of elements of the two-spin reduced density matrix
for the transverse field Ising model with $N=1000$ sites. Inset: 
$d\rho_{22}/d\lambda$ and $d\rho_{44}/d\lambda$ diverge logarithmically
as a function of $N$. They are fitted by $x_{{\textrm{ab}}}\,{\textrm{ln}}\,N + const$, 
with $x_{22}=0.135$ and $x_{44}=0.024$.}
\label{f1}
\end{figure}

Some further features following from this general analysis are: (1) If $
[\partial M(\rho ^{ij})/\partial \lambda ]_{\lambda _{c}}$ diverges then the 
\emph{maximal} entanglement will not occur at the critical point $\lambda
_{c}$. (2) Concerning the behavior in the vicinity of the critical
point: our results above show that 
the speed of divergence of both energy and the entanglement measures is
dominated by the fastest among the $\partial \rho _{ab}^{ij}/\partial
\lambda $ (as illustrated in Fig.~\ref{f1}). 
Therefore $\partial M(\rho ^{ij})/\partial
\lambda $ should have similar divergent properties to the second derivative
of energy. This is indeed the behavior observed for the transverse field Ising model
in Ref.~\cite{Osterloh:02}. (3) Examples exist wherein the max/min
evaluations required by the definition of bipartite entanglement measures generate 
a singularity related to the derivative of these measures, without an
associated QPT~\cite{Yang:04}; condition (a) of our Theorem excludes
such (artificial) singularities. Moreover max/min can also eliminate singularities, 
a possibility which is excluded from consideration through condition~(c). Next we 
consider examples to illustrate our general formalism.

\textit{Frustrated two-leg spin}-$1/2$ \textit{ladder}.--- The Hamiltonian
for this model is 
$H_{\mathrm{ladder}}=\sum_{\left\langle ij\right\rangle }J_{ij}
\overrightarrow{S}_{i}\cdot \overrightarrow{S}_{j}-h\sum
\limits_{i=1}^{N}S_{i}^{z}$,
where $\overrightarrow{S}_{i}$ is the spin operator vector at site $i$, the exchange
interaction along the rungs is $J_{ij}=J_{R}$, and both the intra-chain
nearest-neighbor and diagonal exchange interactions are $J_{ij}=J$. We
further assume $J_R > \gamma J$, with $\gamma \approx 1.401$~\cite{Bose:02}. This
model is exactly solvable and exhibits 1QPTs for $h_{c_{1}}=J_{R}$
and $h_{c_{2}}=J_{R}+2J$. An analysis of pairwise
entanglement for this model can be found in Ref.~\cite{Bose:02}. For
$h<h_{c_{1}}$, and in the limit $N \rightarrow \infty$,
the ground state is a tensor product of (entangled) singlets, $
(|01\rangle -|10\rangle )/\sqrt{2}$, along the rungs. When $
h_{c_{1}}<h<h_{c_{2}}$, the ground state consists of rungs which are
alternately in singlet and (unentangled) $S^{z}=1$ triplet spin
configurations, $|00\rangle $. For $h>h_{c_{2}}$, the ground state is a tensor product of
all rungs in the $S^{z}=1$ triplet state. The density matrix elements 
of the rungs are characterized by the following step-function 
discontinuities at the two critical points: 
\begin{eqnarray}
\rho^{r_i}_{22} &=&\rho^{r_i}_{33}=-\rho^{r_i}_{32}=-\rho^{r_i}_{23}=\left\{ 
\begin{array}{l}
\frac{1}{2},\text{\thinspace \thinspace } h<h_{c_{i}} \\ 
0,\text{\thinspace \thinspace } h \ge h_{c_{i}}
\end{array}
\right.  \nonumber \\
\rho^{r_i}_{11} &=&\left\{ 
\begin{array}{l}
0,\text{\thinspace \thinspace } h<h_{c_{i}} \\ 
1,\text{\thinspace \thinspace \thinspace } h \ge h_{c_{i}}
\end{array}
\right.
\label{rho-spinladder}
\end{eqnarray}
where $r_i$, with $i=1,2$, denotes rungs that transition to the $S^z=1$ configuration 
at the critical point $h_{c_i}$. All other density matrix elements for the rungs vanish. 
The ground state of the system is two-fold degenerate when $h_{c_1}\le h<h_{c_2}$. The  
density operator for a rung is then represented by a statistical mixture of 
the broken-symmetry states $\rho^{r_1}$ and $\rho^{r_2}$, with equal probabilities. Indeed, for a general value 
of $h$, we can write the rung density matrix as $\rho^{r}=(\rho^{r_1}+\rho^{r_2})/2$. 
Below $h_{c_2}$, the ground state energy is given 
by the sum of the energies of each rung, due to the fact that all couplings proportional to $J$ 
vanish when acting on a singlet. Using Eq.~(\ref{erho}) the energy density
can be then written, for $h<h_{c_2}$, as 
${\cal E} = \frac{1}{2} [ \frac{J_{R}}{4} (\rho^{r}_{11}-\rho^{r}_{22}-\rho^{r}_{33}
+\rho^{r}_{44}+2(\rho^{r}_{32}+\rho^{r}_{23}))
- h(\rho^{r}_{11}-\rho^{r}_{44})]$.
For $h\ge h_{c_2}$, contributions of the $J$ sector must be considered in the expression above. 
However the quantity $\partial {\cal E} /\partial h$, which characterizes the 1QPTs 
in this model, can be obtained directly from Eq.~(\ref{fd}) for any $h$, resulting in
$\partial_h \mathcal{E} = -\frac{1}{2}\, \rho^{r}_{11} = 
-\frac{1}{4}(\rho^{r_1}_{11}+\rho^{r_2}_{11})$,
where we have used that $\rho^r_{44}=0$. It then follows from
Eq.~(\ref{rho-spinladder}) that $\partial_h \mathcal{E}$ is discontinuous at both $h_{c_1}$ 
and $h_{c_2}$. The same discontinuous
behavior is immediately revealed in the bipartite entanglement of the
spins sharing a rung. For these pairs a direct calculation shows that
the negativity and 
concurrence (which here turn out to be equal) read
$\mathcal{N}=C=1-\rho^r_{11}=1-\frac{1}{2}(\rho^{r_1}_{11}+\rho^{r_2}_{11})$,
which, therefore, are discontinuous functions at both $h=h_{c_1}$ and $h=h_{c_2}$. 
We thus find the remarkably simple result 
$\partial \mathcal{E}/\partial h=(\mathcal{N}-1)/2$,
which can also be seen as a general consequence of Eq.~(\ref{neg}). This expression exemplifies 
how entanglement directly detects a 1QPT.

\textit{Permutation invariance and the transverse field Ising chain}.--- We
consider now the case of Hamiltonians whose ground states are invariant
under a permutation $P_{ik}$ of an arbitrary pair $(i,k)$ of spins. In this
case $P_{ik}\left\vert \psi \right\rangle =\pm \left\vert \psi \right\rangle 
$ so that $\left\vert \psi \right\rangle \left\langle \psi \right\vert
=P_{ki}P_{lj}\left\vert \psi \right\rangle \left\langle \psi \right\vert
P_{lj}P_{ki}.$ Therefore, from the general expression for the two-spin reduced
density operator ${\hat{\rho}}^{ij}$:
$\left\langle \gamma _{i}\delta _{j}\right\vert \widehat{\rho }
^{ij}\left\vert \alpha _{i}\beta _{j}\right\rangle =\sum_{m}\left\langle
\gamma _{i}\delta _{j}m|\psi \right\rangle \left\langle \psi |m\alpha
_{i}\beta _{j}\right\rangle =  \sum_{m}\left\langle \gamma _{k}\delta _{l}m|\psi \right\rangle
\left\langle \psi |m\alpha _{k}\beta _{l}\right\rangle =\left\langle \gamma
_{k}\delta _{l}\right\vert \widehat{\rho }^{kl}\left\vert \alpha _{k}\beta
_{l}\right\rangle$,
i.e., $\rho ^{ij}=\rho ^{kl}$. If only a constant nearest-neighbor
interaction is taken into account then $\mathbf{U}(i,i+1)=\mathbf{U}(j,j+1)=
\mathbf{U}$ $(\forall \,i,j)$. Then, denoting $\rho _{i,i+1}=\rho
_{j,j+1}=\rho $ $(\forall \,i,j)$, we have 
$\mathcal{E}(\rho )=\mathrm{Tr}(\mathbf{U}\rho)$.
As a specific example, consider the transverse field Ising chain with
constant nearest-neighbor interactions, whose Hamiltonian is 
$H=-J\sum_{i=1}^{N}( \lambda \sigma _{i}^{x}\sigma _{i+1}^{x}+\sigma
_{i}^{z})$,
where $N$ is the number of spins along the chain, $\sigma _{i}^{\alpha }$
are the Pauli operators for a spin at site $i$, and we use periodic boundary
conditions. Setting $J=1$, we obtain from Eq.~(\ref{erho}) that 
$\mathcal{E}(\rho )=-\left\langle \psi \right\vert \sigma ^{z}\left\vert \psi
\right\rangle -2\lambda (\rho _{14}+\rho _{23})$, 
where the site-independent ground state expectation value of $\sigma ^{z}$
is $\left\langle \psi \right\vert \sigma ^{z}\left\vert \psi \right\rangle
=\rho _{11}-\rho _{44}$. This model presents a 2QPT at $\lambda
_{c}=1$ \cite{Sachdev:book}. This can be identified within our framework by 
taking the second derivative of $\mathcal{E}(\rho )$, yielding 
\begin{equation}
\frac{\partial ^{2}\mathcal{E}(\rho )}{\partial \lambda ^{2}}=-2\frac{1}{
\lambda }\frac{\partial }{\partial \lambda }(\rho _{22}+\rho _{44}),
\label{sde}
\end{equation}
where we used  $\sum_{ij}\mathrm{Tr}(\mathbf{U}(ij)\frac{\partial \rho ^{ij}}{\partial
  \lambda })=0$ and $\mathrm{Tr}\rho =1$. We have
calculated the $\rho _{ij}$ using the standard method of fermionization and
a Bogoliubov transformation \cite{Sachdev:book}. At the critical point $\lambda _{c}$, Eq.~(\ref{sde}) displays a divergence in the limit of an
infinite chain. This 2QPT originates from the singular behavior
of $\partial _{\lambda }\rho _{22}$ and $\partial _{\lambda }\rho _{44}$, as
shown in Fig.~\ref{f1}. It is clear from this figure that $\partial _{\lambda }\rho
_{22}$ is dominantly responsible for the divergence. 
\begin{figure}[ht]
\centering {\includegraphics[angle=0,scale=0.29]{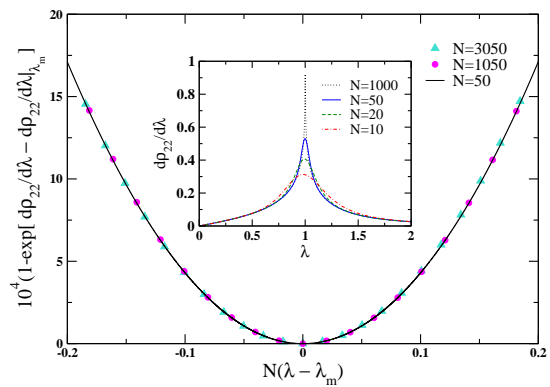}}
\caption{Finite size scaling of $\partial _{\protect\lambda }\protect\rho 
_{22}$ with the number $N$ of sites for the transverse field Ising chain. $
\partial _{\protect\lambda }\protect\rho _{22}$ is a function of $N^{1/
\protect\nu }(\protect\lambda -\protect\lambda _{m})$ only, with the Ising
model critical exponent $\protect\nu =1$, and $\protect\lambda _{m}$ being
the position of the maximum of $\protect\rho _{22}(N,\protect\lambda )$. All
the data from $N=50$ to $N=3050$ collapse onto a single curve. Inset: $
\partial _{\protect\lambda }\protect\rho _{22}$ before scaling, showing
increase in singular peak sharpness with $N$ and shift of $\protect\lambda_{m}$. }
\label{f2}
\end{figure}

Now, concerning the
ground state nearest-neighbor bipartite entanglement, the global $\pi$-rotation 
invariance of the model about the spin $z$-axis ($Z_2$-symmetry) and a detailed computation of the 
density matrix elements leads to  
$C=\mathcal{N}=2(\rho _{41}-\rho _{22})$.
As shown in Ref.~\cite{Syl:03}, the concurrence in this case is not modified by spontaneous symmetry breaking. 
In the limit $N\rightarrow \infty $, $(\partial C/\partial \lambda)|_{\lambda _{c}}$ is 
logarithmically divergent~\cite{Osterloh:02}. This
result is here seen to be a direct consequence of the singular behavior of $
\partial _{\lambda }\rho _{22}$, just as in the second derivative of energy,
since $\partial _{\lambda }\rho _{41}$ is a smooth function of $\lambda $.
Therefore $\partial ^{2}\mathcal{E}/\partial \lambda ^{2}$ and $\partial
C/\partial \lambda $ exhibit similar critical behavior through their
dependence upon $\partial \rho_{22}/\partial \lambda $, whose finite-size scaling is 
shown in Fig.~\ref{f2}. The conclusion of \cite{Osterloh:02}, that the concurrence detects the
phase transition in the Ising model, is thus simply explained within our
framework. 

We gratefully acknowledge financial support
from CNPq-Brazil (to M.S.S.), and the Sloan Foundation, PREA and NSERC
(to D.A.L.). We thank Prof. L.-M. Duan and Dr. P. Zanardi for inspiring
discussions and Dr. M.-F. Yang for useful correspondence.

\end{document}